\documentstyle[amssymb,epsfig,12pt]{article}
\input{tcilatex}

\begin{document}

\bigskip 
\begin{titlepage}
\bigskip \begin{flushright}
hep-th/0203003\\
WATPPHYS-TH02/02
\end{flushright}

%\maketitle

\vspace{1cm}

\begin{center}
{\Large \bf {Vortices in De Sitter Spacetimes}}\\
\end{center}
\vspace{2cm}
\begin{center}
 A.M. Ghezelbash$^{\dagger}${%
\footnote{%
EMail: amasoud@sciborg.uwaterloo.ca}} and R. B. Mann$^\ddagger$
\footnote{
EMail: mann@avatar.uwaterloo.ca}\\
$^{\dagger,\ddagger}$Department of Physics, University of Waterloo, \\
Waterloo, Ontario N2L 3G1, Canada\\
$^{\dagger}$Department of Physics, Alzahra University, \\
Tehran 19834, Iran\\
\vspace{1cm}
%PACS numbers:  
 %11.15.-q, 11.25.Hf, 04.60.-m, 11.10.Lm\\
%\vspace{2cm}
\today\\
\end{center}

\begin{abstract}
We investigate vortex solutions to the Abelian Higgs field equations in 
a four dimensional de Sitter spacetime background. We obtain both static
and dynamic solutions with axial symmetry that are generalizations of the 
Nielsen-Olesen gauge vortices in flat spacetime.  The static solution is
located in the static patch of de Sitter space.  
We numerically solve the field equations in an inflationary (big bang)
patch and find a time dependent vortex soution, whose effect 
to create a deficit angle in the spacetime.
We show that this solution can be interpreted in terms of a renormalization group 
flow in  accord with a generalized {$c$}-theorem, providing evidence in favour of a 
dS/CFT correspondence.
\end{abstract}
\end{titlepage}\onecolumn

\begin{center}
\bigskip %%\pacs{04.60.-m,04.62.+v,04.70.-s,04.70.dy,11.25.-w}
\end{center}

\section{Introduction}

De Sitter spacetimes are becoming of increasing interest in theoretical
physics for a variety of reasons. They provide interesting arenas in which
to study the classical no-hair conjecture first proposed by Ruffini and
Wheeler \cite{Ruf}, which states that after a given distribution of matter
undergoes complete gravitational collapse, the only long range information
of the resultant black hole is its electromagnetic charge, mass and angular
momentum. The verification of this conjecture for a scalar field minimally
coupled to gravity in asymptotically flat black-hole spacetimes has been
extended to its de Sitter counterpart \cite{Sud},\cite{Torii1}. While it is
tempting to extend the no-hair theorem claim to all forms of matter, it is
known that some long range Yang-Mills and/or quantum hair can be ``painted''
on a black hole \cite{Eli}. Explicit calculations have been carried out
which verify the existence of a long range Nielsen-Olesen vortex solution as
a single stable hair for a Schwarzschild black hole in four dimensions \cite
{Achu}, although it might be argued that this situation falls outside the
scope of the classical no-hair theorem due to the non trivial topology of
the string configuration.

Another motivation for studying de Sitter (dS) spacetimes is connected with
the recently proposed holographic duality between quantum gravity on dS
spacetime and a quantum field theory living on the past boundary of dS
spacetime \cite{Stro}. This proposed correspondence is undergoing extensive
study in various directions \cite{Noj}-\cite{Leb}. So far this work suggests
that the conjectured dS/CFT correspondence has a lot of similarity with the
AdS/CFT correspondence, although some interpretive issues remain \cite{GH}.
\ 

Motivated by these considerations, we pursue in this paper a study of vortex
solutions in de Sitter spacetimes. We have already shown that the $U(1)$
Higgs field equations have a vortex solution in both a four dimensional AdS
spacetime \cite{Deh} and in a four dimensional Schwarzschild-AdS background 
\cite{Deh2}. Employing the well known AdS/CFT correspondence conjecture, the
boundary conformal field theory can detect the presence of the vortex in the
four dimensional AdS spacetime: the mass density of the vortex solution is
encoded in the discontinuity of the two-point correlation function of the
dual conformal operator \cite{Deh}. We have also shown that vortex solutions
exist in the background of rotating Kerr-AdS and charged
Reissner-Nordstrom-AdS black holes \cite{R}, even in the extremal case.\
Concurrently, it has recently been shown that in an asymptotically AdS
spacetime that a black hole can have scalar hair \cite{Torii2}. Indeed,
insofar as the no-hair theorem is concerned it has been shown that there
exists a solution to the $SU(2)$ Einstein-Yang-Mills equations which
describes a stable Yang-Mills hairy black hole that is asymptotically AdS 
\cite{Eli}.

We therefore seek to learn if an analog of vortex holography discussed in 
\cite{Deh}\thinspace holds true for dS spacetime. In this article we take
the first steps toward consideration of such a holographic phenomenon by
searching for possible solutions of the Abelian Higgs field equations in a
four dimensional dS background. Although an analytic solution to these
equations appears to be intractable, we confirm by numerical calculation
that vortex solutions do exist in dS spacetime. We find both static and
time-dependent vortex solutions. The static solution corresponds to a core
of vortex field energy located within the cosmological horizon in a static
patch of dS spacetime. In the time-dependent solution, the energy of the
vortex core is diluted by the cosmological expansion in an inflationary
patch. To our knowledge this is the first construction of vortex solutions
in asymptotically dS spacetimes.

We also consider the implications of our solutions for the recently
conjectured dS/CFT correspondence \cite{Stro}. \ We compute the
renormalization group flow associated with the time-dependent vortex
solution in the inflationary patch and find that the generalized $c$%
-function monotonically increases, in accord with the generalized $c$%
-theorem \cite{Leb}.

In section two, we solve numerically the Abelian-Higgs equations in the
static dS background for different values of the cosmological constant. In
section three, we compute the effect of the vortex solution on the dS
spacetime. In section four, we solve numerically the same equations in a big
bang patch of dS background. This is the first investigation of time
dependent vortices in curved spacetime. In section five, we obtain the
behaviour of the dS $c$-function of this solution and find that it increases
(decreases) in an expanding (contracting) dS patch. We give some closing
remarks in the final section.

\section{Abelian Higgs Vortex in static de Sitter Spacetime}

In this section, we consider the abelian Higgs Lagrangian in the background
of de Sitter spacetime 
\begin{equation}
{\cal L}(\Phi ,A_{\mu })=-\frac{1}{2}({\cal D}_{\mu }\Phi )^{\dagger }{\cal D%
}^{\mu }\Phi -\frac{1}{16\pi }{\cal F}_{\mu \nu }{\cal F}^{\mu \nu }-\xi
(\Phi ^{\dagger }\Phi -\eta ^{2})^{2}  \label{Lag}
\end{equation}
where $\Phi $ is a complex \ scalar Klein-Gordon field, ${\cal F}_{\mu \nu }$
is the field strength of \ the electromagnetic field $A_{\mu }$ and ${\cal D}%
_{\mu }=\nabla _{\mu }+ieA_{\mu }$ in which $\nabla _{\mu }$ is the
covariant derivative. We employ Planck units $G=\hbar =c=1$ which implies
that the Planck mass is equal to unity. We use the following four
dimensional static dS spacetime background 
\begin{equation}
ds^{2}=-(1-\frac{r^{2}}{l^{2}})dt^{2}+\frac{1}{(1-\frac{r^{2}}{l^{2}})}%
dr^{2}+r^{2}(d\theta ^{2}+\sin ^{2}\theta \,d\varphi ^{2})  \label{dsmetric}
\end{equation}
where the cosmological constant $\Lambda $ is equal to $\frac{3}{l^{2}}.$
The horizon is at $r=l$ and so the range of the \ coordinate $r$ is bounded
to $0\leq r\leq l$. In this coordinate system $\partial /\partial t$ is a
future-pointing timelike Killing vector in only one diamond of the Penrose
diagram \cite{VOLO} which generates the symmetry $t\rightarrow t+t_{0}$ for
any constant $t_{0}.$ In other diamonds of the Penrose diagram, this Killing
vector is spacelike or else past-pointing timelike. After redefining the new
fields $X(x^{\mu }),P_{\mu }(x^{\nu })$ by 
\begin{equation}
\begin{array}{c}
\Phi (x^{\mu })=\eta X(x^{\mu })e^{i\omega (x^{\mu })} \\ 
A_{\mu }(x^{\nu })=\frac{1}{e}(P_{\mu }(x^{\nu })-\nabla _{\mu }\omega
(x^{\mu }))
\end{array}
\label{new}
\end{equation}
and employing a suitable gauge, the equations of motion for a string with
winding number $N$ are 
\begin{equation}
(1-\frac{\rho ^{2}}{l^{2}})\frac{d^{2}X}{d\rho ^{2}}+(\frac{1}{\rho }-\frac{%
4\rho }{l^{2}})\frac{dX}{d\rho }-\frac{1}{2}X(X^{2}-1)-\frac{N^{2}}{\rho ^{2}%
}XP^{2}=0  \label{eqxds}
\end{equation}
\begin{equation}
(1-\frac{\rho ^{2}}{l^{2}})\frac{d^{2}P}{d\rho ^{2}}-\frac{dP}{d\rho }(\frac{%
1}{\rho }+\frac{2\rho }{l^{2}})-\alpha PX^{2}=0  \label{eqpds}
\end{equation}
where $\rho =r\sin \theta $ and $\alpha =\frac{4\pi e^{2}}{\xi }.$ Note that
by changing $l$ to $il$, equations (\ref{eqxds}) and (\ref{eqpds}) change to
the equations of motion in an AdS background \cite{Deh}, or to the $\ m=0$
equations discussed in \cite{Deh2}. We have solved the above equations
numerically for dS spacetimes with $l=3,5,10$ and unit winding number using
the over-relaxation method \cite{num}. 
\begin{figure}[tbp]
\begin{center}
\epsfig{file=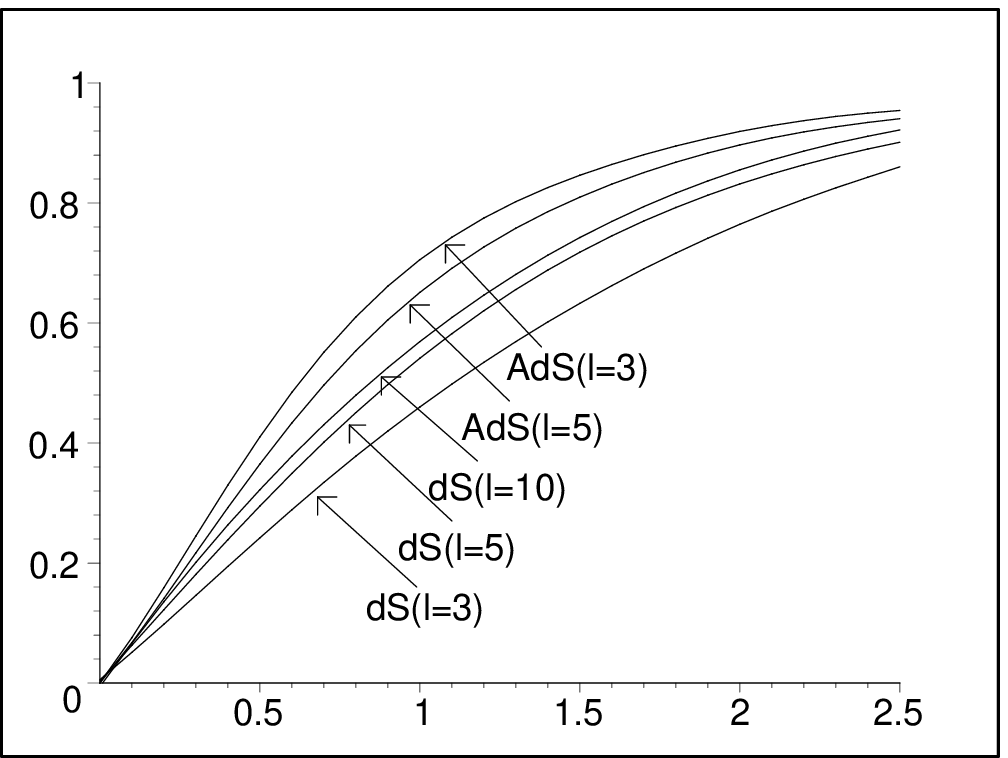,width=0.4\linewidth} %
\epsfig{file=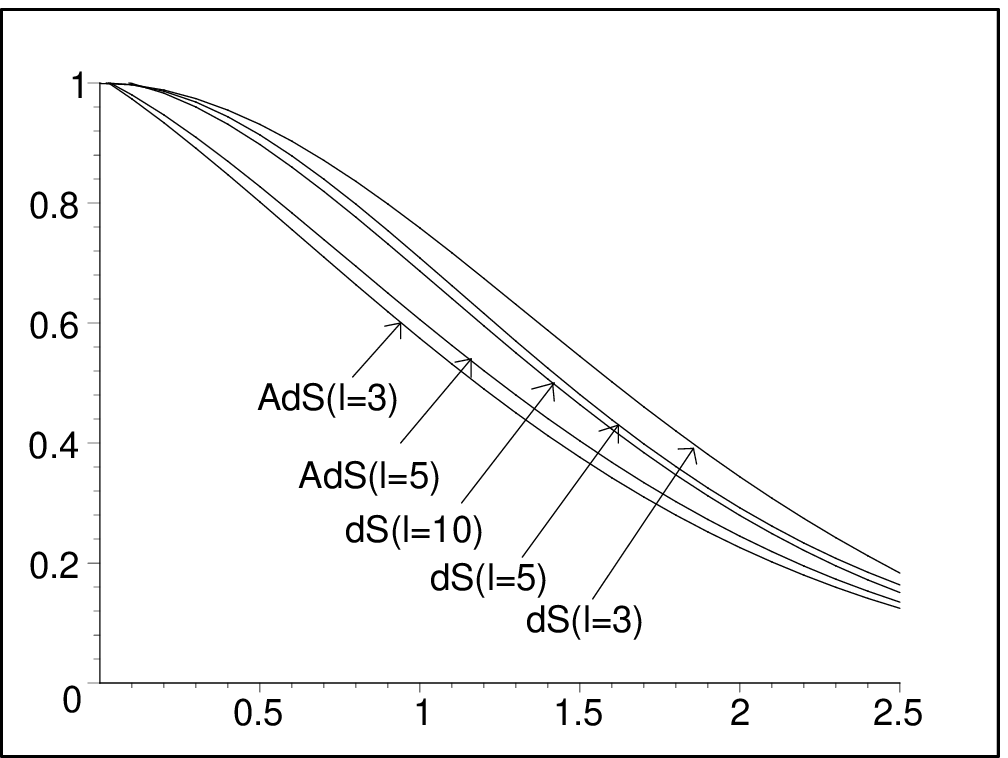,width=0.4\linewidth}
\end{center}
\caption{{}$X(\protect\rho )$ and $P(\protect\rho )$ fields of a vortex, in
dS spacetime for the different values of $l=3,5,10$. These fields in the AdS
spacetime with $l=3,5$ have similar behaviour. }
\label{fig12}
\end{figure}

Figure \ref{fig12} shows the results for our calculation and also compares
them with the vortex fields in the AdS spacetime \cite{Deh}. For fixed $\rho
\leq 3$, the $X$ \ field of the dS spacetime decreases with decreasing $l$,
in contrast to the AdS spacetime which $X$ field increases with decreasing $%
l.$

In general the value of the $X$ field for AdS spacetime is always greater
than its value for any dS spacetime for a given $\rho $. In other words, by
increasing the cosmological constant $\Lambda $ from $-\infty $ to $+\infty ,
$ the $X$ field decreases for fixed $\rho $. The $X$ field of flat spacetime
is located between the dS and AdS cases where $\left| l\right| =\infty $.
Over this same range of $\Lambda $ the value of the $P$ field increases.{\bf %
\ }Physically the negative cosmological constant exerts additional pressure
on the vortex, causing it to become thinner; a positive cosmological
constant has the opposite effect, causing the core of the vortex to expand
beyond the thickness it would have in flat spacetime.\ Moreover, we expect
that the effect of the vortex on dS spacetime is to create a deficit angle
in the metric (\ref{dsmetric}) by replacing $\varphi \rightarrow \alpha
\varphi ,$ which $\alpha $ is a constant. In the next section, we verify
this point.

For $r\geq l,$ the coefficients of the $\frac{d^{2}X}{d\rho ^{2}}$ and $%
\frac{d^{2}P}{d\rho ^{2}}$ terms in equations (\ref{eqxds}) and (\ref{eqpds}%
) change sign and the numerical solution of the corresponding equations
outside the cosmological horizon shows that the values of the $X$ and $P$
fields remain at their respective constant values of $1$ and $0$.
Consequently the vortex is confined totally behind the cosmological horizon.
In the next section we will find that in a different patch of dS spacetime
the vortex solution exists everywhere in space and changes with time.

\section{Vortex self gravity in static de Sitter space time}

In this section, we consider the effect of \ the vortex on dS$_{4}$
spacetime described by the static metric (\ref{dsmetric}) . This requires
finding the solutions of the coupled Einstein-Abelian Higgs differential
equations in dS$_{4}$. This is a formidable problem -- even for flat
spacetime no exact solutions have yet been found. Using a thin-core
approximation and numerical methods, we obtained in \cite{Deh2} the effect
of the vortex on AdS$_{4}$. Here, we use the same method to obtain the
effect of the vortex on dS$_{4}$ spacetime.

\begin{figure}[tbp]
\begin{center}
\epsfig{file=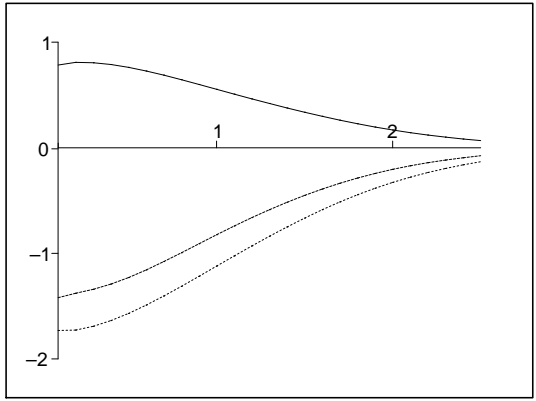,width=0.4\linewidth}
\end{center}
\caption{$T_{\protect\varphi }^{\protect\varphi (0)}$(solid),$T_{t}^{t(0)}$%
(dash),$T_{r}^{r(0)}+T_{\protect\theta }^{\protect\theta (0)}$ (dash-dotted)
curves of a vortex in dS spacetime with $l=3,$ versus $\protect\rho .$ }
\label{fig125}
\end{figure}

To obtain physical results, we make some approximations. First, we again
assume the thin-core approximation, namely that the thickness of the vortex
is much smaller that all other relevant length scales. Second, we assume
that the gravitational effects of the string are weak enough so that the
linearized Einstein-Abelian Higgs differential equations are applicable. For
convenience, in this section we use the following form of the metric of dS$%
_{4}:$

\begin{equation}
ds^{2}=-\widetilde{{\cal A}}(r,\theta )^{2}dt^{2}+\widetilde{{\cal B}}%
(r,\theta )^{2}d\varphi ^{2}+\widetilde{{\cal C}}(r,\theta )(\frac{dr^{2}}{1-%
\frac{r^{2}}{l^{2}}}+r^{2}d\theta ^{2})  \label{ABCmetric}
\end{equation}
In the absence of the vortex, we must have ${\cal A}(r,\theta )=\sqrt{1-%
\frac{r^{2}}{l^{2}}}={\cal A}_{0}(r,\theta ),{\cal B}(r,\theta )=r\sin
\theta ={\cal B}_{0}(r,\theta ),{\cal C}(r,\theta )=1$ $={\cal C}%
_{0}(r,\theta )$, yielding the well known metric (\ref{dsmetric}) of pure dS$%
_{4}$. Employing the two assumptions concerning the thickness of the vortex
core and its weak gravitational field, we solve numerically the Einstein
field equations, 
\begin{equation}
G_{\mu \nu }+\frac{3}{l^{2}}g_{\mu \nu }=-8\pi G{\cal T}_{\mu \nu }
\label{Einstein}
\end{equation}
to first order in{\Large \ }$\varepsilon =-8\pi G,$ where ${\cal T}_{\mu \nu
}$ is the energy-momentum tensor of the Abelian Higgs field in the dS
background. By taking $g_{\mu \nu }\simeq g_{\mu \nu }^{(0)}+g_{\mu \nu
}^{(1)}$ , where $g_{\mu \nu }^{(0)}$ is the usual dS$_{4\text{ }}$ metric
and $\ g_{\mu \nu }^{(1)}$ is its first order correction, and writing 
\begin{equation}
\begin{tabular}{l}
$\widetilde{{\cal A}}(r,\theta )={\cal A}_{0}(r,\theta )(1+\varepsilon {\cal %
A}(r,\theta ))$ \\ 
$\widetilde{{\cal B}}(r,\theta )={\cal B}_{0}(r,\theta )(1+\varepsilon {\cal %
B}(r,\theta ))$ \\ 
$\widetilde{{\cal C}}(r,\theta )={\cal C}_{0}(r,\theta )(1+\varepsilon {\cal %
C}(r,\theta ))$%
\end{tabular}
\label{ABCexpa}
\end{equation}
we obtain first-order corrections to the three functions ${\cal A}%
_{0}(r,\theta ),{\cal B}_{0}(r,\theta )$ and ${\cal C}_{0}(r,\theta )$ in (%
\ref{ABCexpa}). Hence in the first approximation the equations (\ref
{Einstein}) become

\begin{equation}
G_{\mu \nu }^{(1)}+\frac{3}{l^{2}}g_{\mu \nu }^{(1)}={\cal T}_{\mu \nu
}^{(0)}  \label{Eineq}
\end{equation}
where ${\cal T}_{\mu \nu }^{(0)}$ is the energy momentum tensor of the
vortex field in the dS$_{4\text{ }}$background metric, and $G_{\mu \nu
}^{(1)}$ is the correction to the Einstein tensor due to $g_{\mu \nu }^{(1)}$%
. The rescaled components of the energy momentum tensor of string in the
background of dS$_{4}$ are given by 
\begin{equation}
\begin{tabular}{l}
$T_{t}^{t(0)}(\rho )=-\frac{1}{2}(\frac{dX}{d\rho })^{2}(1-\frac{\rho ^{2}}{%
l^{2}})-\frac{1}{2}\frac{1}{\rho ^{2}}(\frac{dP}{d\rho })^{2}(1-\frac{\rho
^{2}}{l^{2}})-\frac{1}{2}\frac{P^{2}X^{2}}{\rho ^{2}}-(X^{2}-1)^{2}$ \\ 
$T_{\varphi }^{\varphi (0)}(\rho )=-\frac{1}{2}(\frac{dX}{d\rho })^{2}(1-%
\frac{\rho ^{2}}{l^{2}})+\frac{1}{2}\frac{1}{\rho ^{2}}(\frac{dP}{d\rho }%
)^{2}(1-\frac{\rho ^{2}}{l^{2}})+\frac{1}{2}\frac{P^{2}X^{2}}{\rho ^{2}}%
-(X^{2}-1)^{2}$ \\ 
$(T_{r}^{r(0)}+T_{\theta }^{\theta (0)})(\rho )=-\frac{P^{2}X^{2}}{\rho ^{2}}%
-2(X^{2}-1)^{2}$%
\end{tabular}
\label{stresssph}
\end{equation}
where $T_{\mu }^{\mu (0)}=\xi ^{-1}\eta ^{-4}{\cal T}_{\mu }^{\mu (0)}.$ The
Einstein equations (\ref{Eineq}) are 
\begin{equation}
\begin{tabular}{l}
$(1-\frac{\rho ^{2}}{l^{2}})\frac{d^{2}{\cal B}}{d\rho ^{2}}+2\frac{d{\cal B}%
}{d\rho }(\frac{1}{\rho }-\frac{2\rho }{l^{2}})+\frac{1}{2}(1-\frac{\rho ^{2}%
}{l^{2}})\frac{d^{2}{\cal C}}{d\rho ^{2}}-\frac{\rho }{l^{2}}\frac{d{\cal C}%
}{d\rho }+\frac{3{\cal C}}{l^{2}}=T_{t}^{t(0)}$ \\ 
$(1-\frac{\rho ^{2}}{l^{2}})\frac{d^{2}{\cal A}}{d\rho ^{2}}-\frac{4\rho }{%
l^{2}}\frac{d{\cal A}}{d\rho }+\frac{1}{2}(1-\frac{\rho ^{2}}{l^{2}})\frac{%
d^{2}{\cal C}}{d\rho ^{2}}-\frac{\rho }{l^{2}}\frac{d{\cal C}}{d\rho }+\frac{%
3{\cal C}}{l^{2}}=T_{\varphi }^{\varphi (0)}$ \\ 
$(1-\frac{\rho ^{2}}{l^{2}})(\frac{d^{2}{\cal A}}{d\rho ^{2}}+\frac{d^{2}%
{\cal B}}{d\rho ^{2}})+\frac{2}{\rho }(\frac{d{\cal A}}{d\rho }+\frac{d{\cal %
B}}{d\rho })(1-3\frac{\rho ^{2}}{l^{2}})+\frac{6{\cal C}}{l^{2}}%
=T_{r}^{r(0)}+T_{\theta }^{\theta (0)}$%
\end{tabular}
\label{eins}
\end{equation}
We consider the case $l=3,$ for which the behaviour of $X$ and $P$ fields
are given in figure \ref{fig12}. By using equations (\ref{stresssph}), we
obtain figure \ref{fig125} showing the behaviour of the stress tensor
components inside the cosmological horizon as a function of $\rho ,$ whose
minimum value is $0.1.$

{\bf \bigskip }

\begin{figure}[tbp]
\begin{center}
\epsfig{file=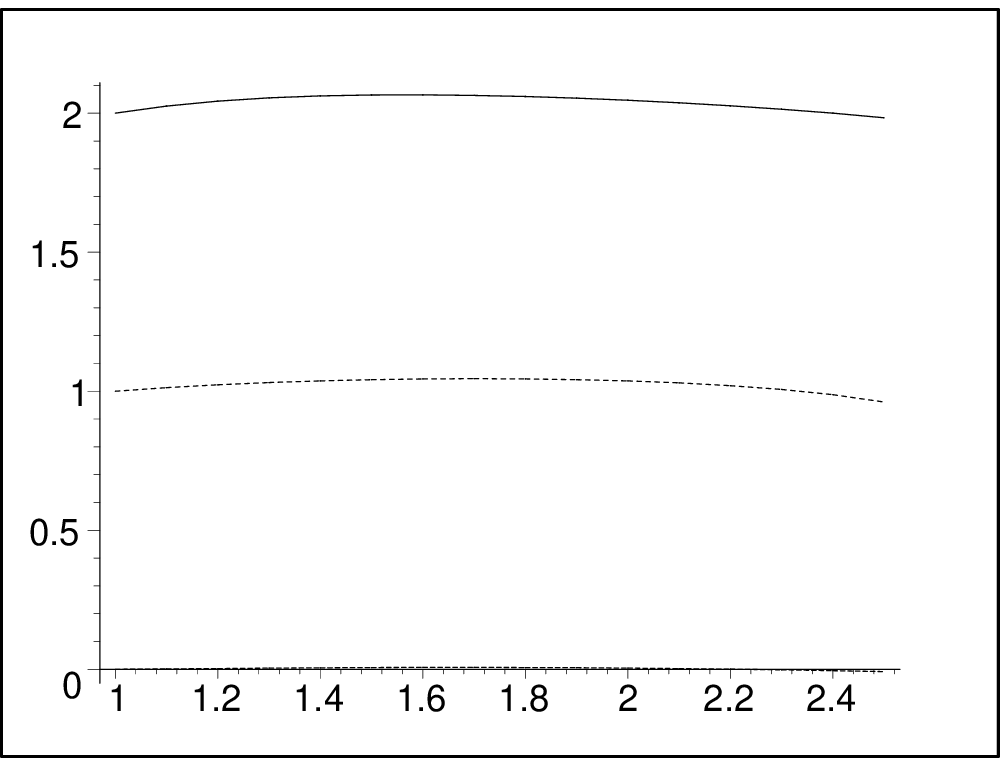,width=0.4\linewidth}%
\epsfig{file=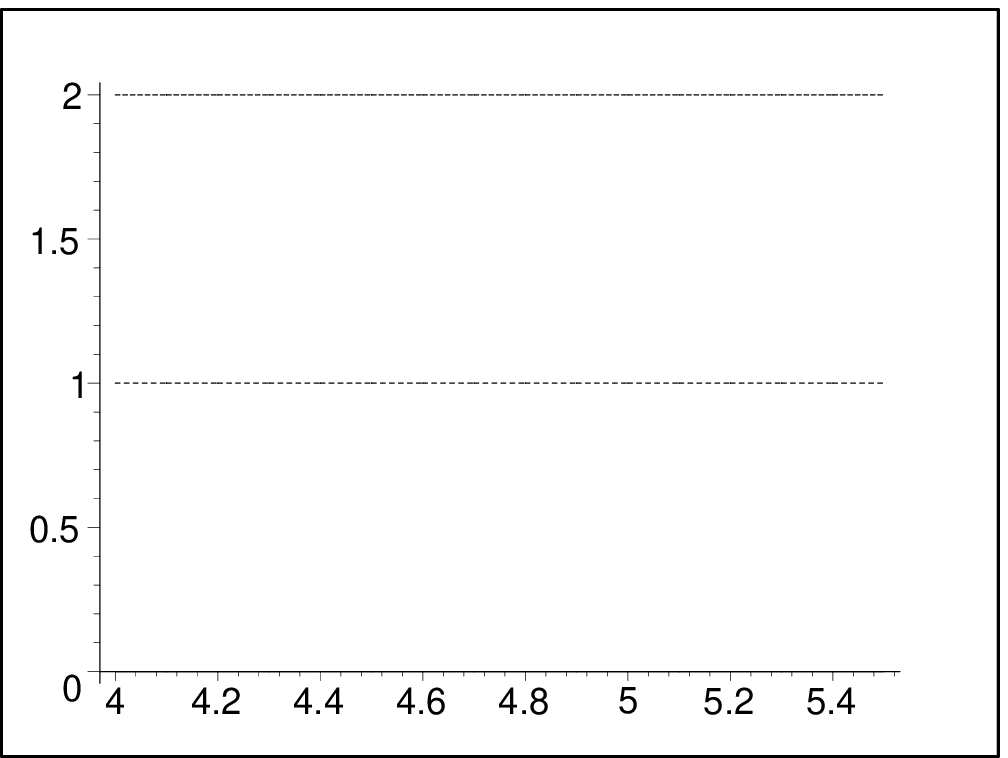,width=0.4\linewidth}
\end{center}
\caption{${\cal A}$(dotted),${\cal B}$(solid)$,{\cal C}$(dashed) versus $%
\protect\rho $, inside and outside the cosmological horizon of the dS
spacetime with $l=3.$ }
\label{fig126}
\end{figure}

We note that outside the cosmological horizon, the stress tensor vanishes
due to the constant values of the vortex fields. With this knowledge, we
solve the coupled differential equations (\ref{eins}), inside and outside
the cosmological horizon which gives the behaviour of the functions ${\cal A}%
(\rho ),{\cal B}(\rho )$ and ${\cal C}(\rho ).$ The results are plotted in
figure \ref{fig126}. These results emphasize that the functions ${\cal A}%
(\rho )\simeq 1,{\cal B}(\rho )\simeq 2$ and ${\cal C}(\rho )\simeq 0$ are
constant over the entire range of $\rho $ to this order in $\varepsilon $. A
very tiny deviation from the constant values in the curves ${\cal A}(\rho ),%
{\cal B}(\rho )$ are due to the numerical calculation. 

Hence by a redefinition of the time coordinate in (\ref{ABCmetric}) the
metric can be rewritten as 
\begin{equation}
ds^{2}=-(1-\frac{r^{2}}{l^{2}})dt^{2}+\frac{1}{(1-\frac{r^{2}}{l^{2}})}%
dr^{2}+r^{2}(d\theta ^{2}+\alpha ^{2}\sin ^{2}\theta \,d\varphi ^{2})
\label{dsdef}
\end{equation}
which is the metric of dS space with a deficit angle. So, the effect of the
vortex on dS$_{4}$ spacetime is to create a deficit angle in the metric (\ref
{dsmetric}) by replacing $\varphi \rightarrow \alpha \varphi ,$ where $%
\alpha \simeq 1+2\varepsilon $ is a constant, since{\Large \ }$\varepsilon
<0.$ The above calculation for the effect of the vortex on dS spacetimes is
not restricted to the special case $l=3$, with other values of cosmological
parameter $l$, the final result is the metric (\ref{dsdef}), with some other
deficit angle $\alpha $.

\section{Abelian Higgs Vortex in a Big Bang Patch of de Sitter Spacetime}

In this section, we consider the abelian Higgs Lagrangian (\ref{Lag}) in the
following coordinate system 
\begin{equation}
ds^{2}=-d\tau ^{2}+e^{2\tau /l}(dx^{2}+dy^{2}+\,dz^{2})  \label{bigbang}
\end{equation}
where the coordinate $\tau \in (-\infty ,+\infty ).$ After using the new
fields $X(\tau ,r,\theta ),P_{\mu }(\tau ,r,\theta )$ given by (\ref{new}),
and applying the radial variable $R=\sqrt{x^{2}+y^{2}}=r\sin \theta ,$ the
equations of motion for a string with winding number $N$ become 
\begin{equation}
e^{-2\tau /l}\frac{\partial ^{2}X}{\partial R^{2}}+\frac{e^{-2\tau /l}}{R}%
\frac{\partial X}{\partial R}-\frac{\partial ^{2}X}{\partial \tau ^{2}}-%
\frac{3}{l}\frac{\partial X}{\partial \tau }-\frac{N^{2}}{R^{2}}XP^{2}-\frac{%
1}{2}X(X^{2}-1)=0  \label{eqxdscrunch}
\end{equation}
\begin{equation}
e^{-2\tau /l}\frac{\partial ^{2}P}{\partial R^{2}}-\frac{e^{-2\tau /l}}{R}%
\frac{\partial P}{\partial R}-\frac{\partial ^{2}P}{\partial \tau ^{2}}-%
\frac{1}{l}\frac{\partial P}{\partial \tau }-\alpha PX^{2}=0
\label{eqpdscrunch}
\end{equation}

By numerically solving equations (\ref{eqxdscrunch},\ref{eqpdscrunch}), we
are able to show that a vortex solution exists on a dS spacetime background
for different values of the winding number $N$ and cosmological parameter $l$%
. As in the pure AdS case \cite{Deh} and asymptotically AdS spacetimes \cite
{Deh2}, \cite{R} the results indicate that increasing the winding number
yields a greater vortex thickness. Furthermore, for a vortex with definite
winding number, the string core increases with decreasing $l$ at constant
positive time. The $X$ \ and $P$ fields less rapidly approach their
respective maximum and minimum values at larger distances as $l$ decreases.
\ 

To obtain numerical solutions of (\ref{eqxdscrunch}) and (\ref{eqpdscrunch}%
), we must first select appropriate boundary conditions. At large distances
physical considerations motivate a clear choice. Since in the limit $%
l\rightarrow \infty $, the results must be in agreement with the results of
flat spacetime, we demand that the solutions approach the solutions of the
vortex equations in the flat spacetime given in ref. \cite{Deh2}. This means
that we demand $X\rightarrow 1$ and $P\rightarrow 0$ as $R$ goes to
infinity. On the symmetry axis of the string, we take $X\rightarrow 0$ and $%
P\rightarrow 1$ for all time. Finally we take $X=1$ and $P=0,$ everywhere
(except on the symmetry axis) at the initial time $\tau =-\infty $. In the
following, we consider the case $N=1.$ We can straightforwardly obtain
similar results for other values of the winding number $N.$

\begin{figure}[tbp]
\begin{center}
\epsfig{file=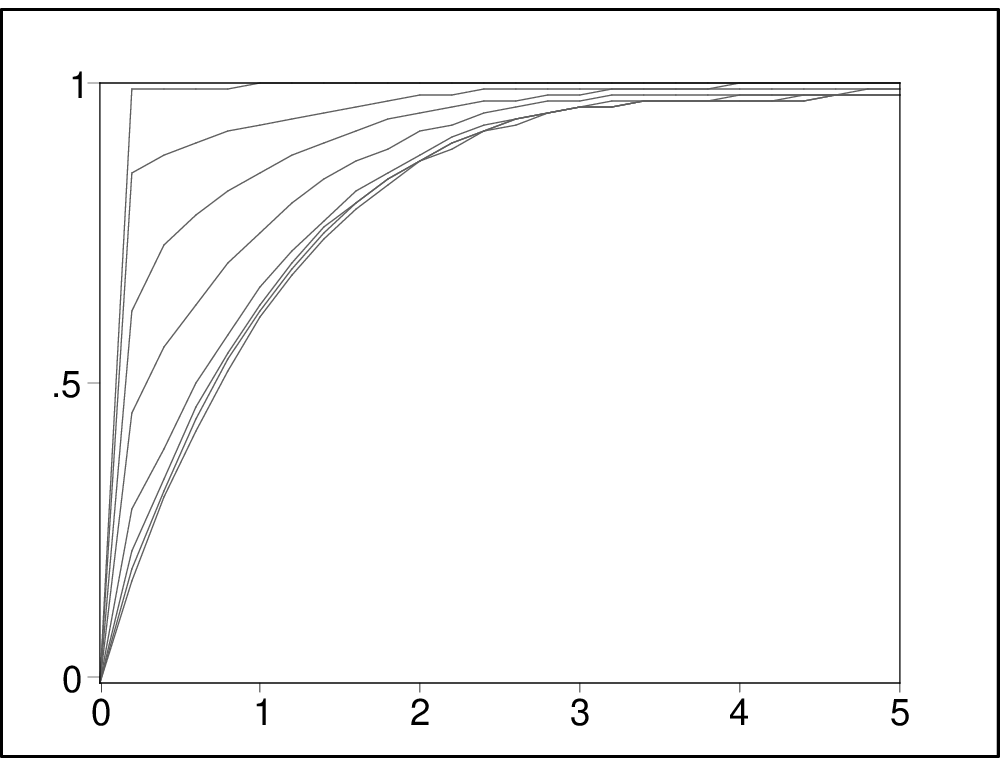,width=0.4\linewidth} %
\epsfig{file=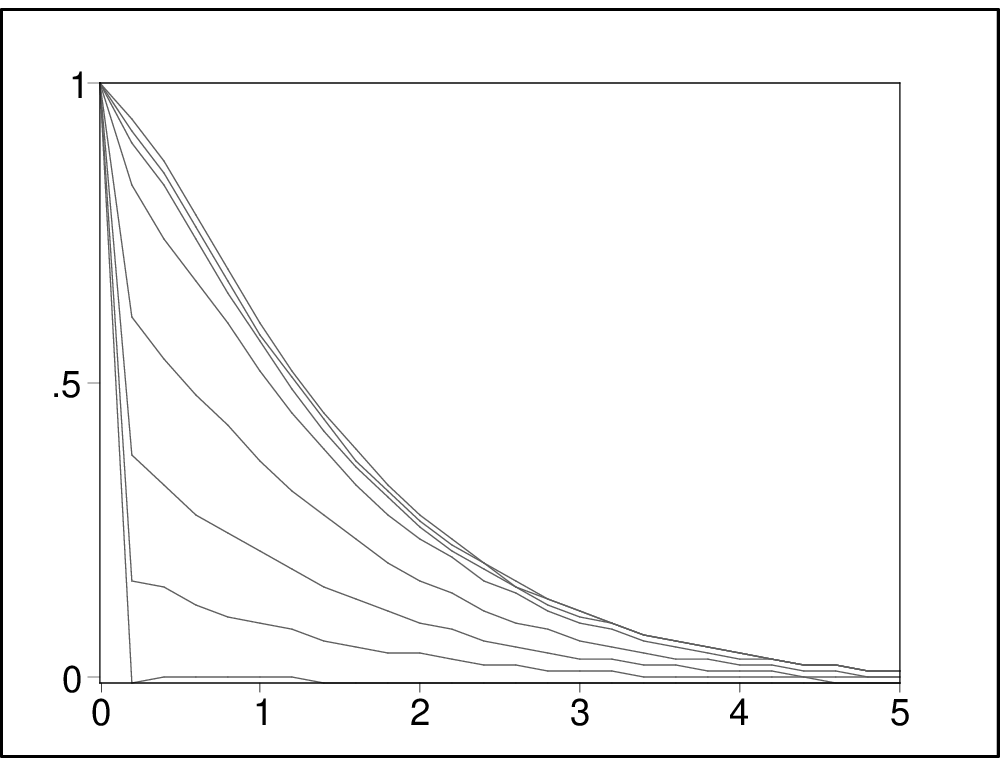,width=0.4\linewidth}
\end{center}
\caption{{}$X(R)$ and $P(R)$ fields of a vortex in dS spacetime with $l=3,$
for the different values of time. Time goes from $-\protect\tau _{\infty }$
to $0,$ from left to right on the curves.}
\label{fig13}
\end{figure}

\begin{figure}[tbp]
\begin{center}
\epsfig{file=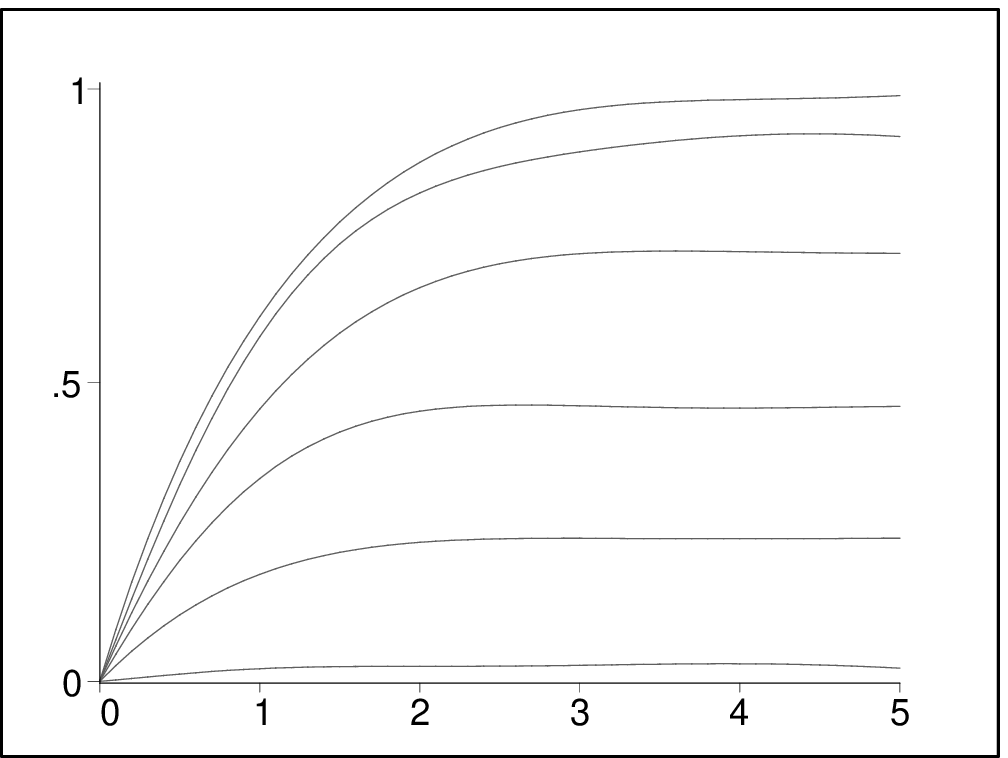,width=0.4\linewidth} %
\epsfig{file=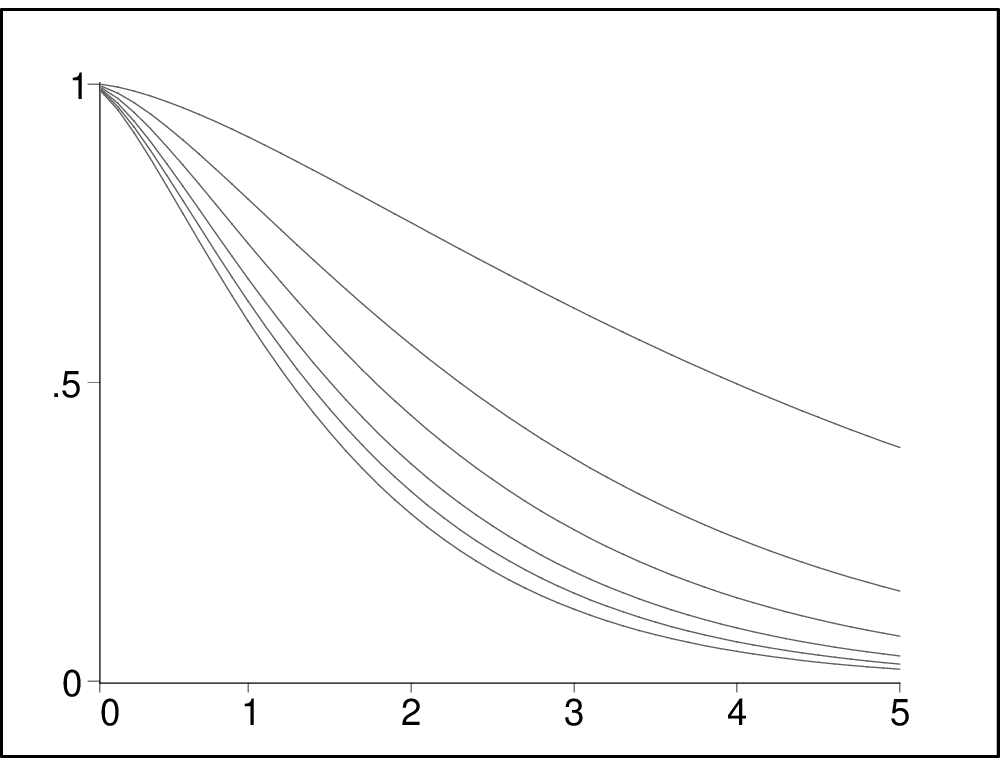,width=0.4\linewidth}
\end{center}
\caption{{}$X(R)$ and $P(R)$ fields of a vortex in dS spacetime with $l=3,$
for the different values of time. Time goes from $0$ to $\protect\tau %
_{\infty },$ from left to right on the curves.}
\label{fig14}
\end{figure}

We then employ a polar grid of points $(R_{i},\tau _{j}),$ where $R$ goes
from $0$ to some large value $R_{\infty }$ which is much greater than $l$
and $\tau $ runs from a large negative number $-\tau _{\infty }$ to a large
positive number $\tau _{\infty }.$ \ Using finite difference methods, we
rewrite the non linear partial differential equation (\ref{eqxdscrunch}) and
(\ref{eqpdscrunch}) as 
\begin{equation}
A_{ij}X_{i+1,j}+B_{ij}X_{i-1,j}+C_{ij}X_{i,j+1}+D_{ij}X_{i,j-1}+E_{ij}X_{i,j}=F_{ij}
\label{findiffxx}
\end{equation}
\begin{equation}
A_{ij}^{\prime }P_{i+1,j}+B_{ij}^{\prime }P_{i-1,j}+C_{ij}^{\prime
}P_{i,j+1}+D_{ij}^{\prime }P_{i,j-1}+E_{ij}^{\prime }P_{i,j}=F_{ij}^{\prime }
\label{findiffxp}
\end{equation}
where $X_{ij}=X(R_{i},\tau _{j})$ and $P_{ij}=P(R_{i},\tau _{j}).$ For the
interior grid points, the coefficients $A_{ij},...,F_{ij\text{ }}^{\prime }$
can be straightforwardly determined from the corresponding continued
differential equations (\ref{eqxdscrunch}) and (\ref{eqpdscrunch}). Using
the well known successive overrelaxation method \cite{num} for the above
mentioned finite difference equations, we obtain the values of $X$ and $P$
fields inside the grid. Initially, inside the grid points, we set the value
of $X$ and $P$ fields $0$ and $1$ respectively which we denote them by $%
X^{(0)}$ and $P^{(0)}$. \ Then these values of $X$ and $P$ fields are used
in the next step to obtain the values of $X$ and $P$ fields inside the grid
which could be denoted by $X^{(1)}$ and $P^{(1)}$. Repeating this procedure,
the value of the each field in the $n$-th iteration is related to the $(n-1)$%
-th iteration by

\begin{equation}
X_{ij}^{(n)}=X_{ij}^{(n-1)}-\omega \frac{\zeta _{ij}^{(n-1)}}{E_{ij}^{(n-1)}}
\label{sor}
\end{equation}
\begin{equation}
P_{ij}^{(n)}=P_{ij}^{(n-1)}-\omega \frac{\varsigma _{ij}^{(n-1)}}{%
E_{ij}^{\prime (n-1)}}  \label{sorp}
\end{equation}
where the residual matrices $\zeta _{ij}^{(n)}$ and $\varsigma _{ij}^{(n)}$
are the differences between the left and right hand sides of the equations (%
\label{findiffx}\ref{findiffxx}) and (\ref{findiffxp}) respectively,
evaluated in the $n$-th iteration. $\omega $ is the overrelaxation
parameter. The iteration is performed many times to some value $n=K,$ such
that $\sum_{i,j}\left| X_{ij}^{(K)}-X_{ij}^{(K-1)}\right| <\varepsilon $ and 
$\sum_{i,j}\left| P_{ij}^{(K)}-P_{ij}^{(K-1)}\right| <\varepsilon $ for a
given error $\varepsilon $. It is a matter of trial and error to find the
value of $\omega $ that yields the most rapid convergence. Some typical
results of this calculation are displayed in figures \ref{fig13} and \ref
{fig14} for value of $\ l=3$. 

In the figure \ref{fig13}, time is running from $-\tau _{\infty }$ to $0$\
whereas in figure \ref{fig14} the time runs from $0$ to $\tau _{\infty }.$
We notice that by increasing the time from $-\tau _{\infty }$ to $0,$ the
string core increases rapidly to a well defined value at time $\tau =0$.
Then increasing the time to more positive values, the string core increases
\ more and more such that at time $\tau =\tau _{\infty },$ the $X$ and $P$
fields approach the constant values $0$ and $1$ respectively. \ So, we
conclude that by increasing the time from $-\tau _{\infty \text{ }}$to $\tau
_{\infty },$ for which the constant time slices of the spacetime (\ref
{bigbang}) become bigger and bigger, the string thickness becomes wider and
wider. Moreover the energy density of the string spreads to bigger and
bigger distances as time increases due to the larger size of the\ constant
time slices.

\begin{figure}[tbp]
\begin{center}
\epsfig{file=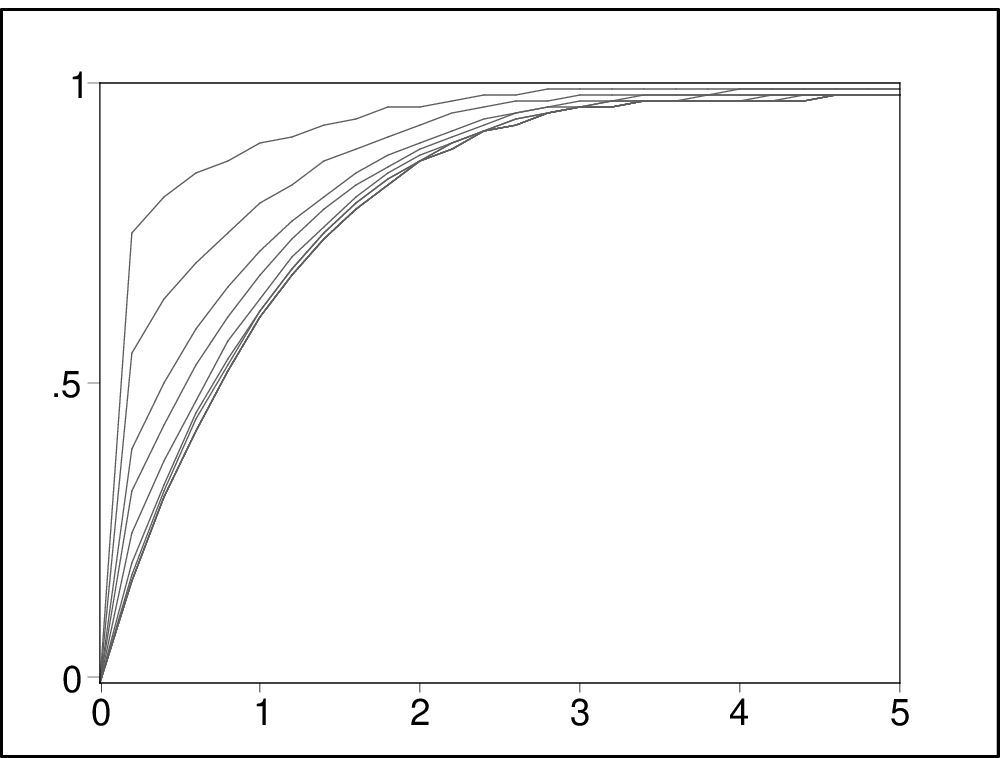,width=0.4\linewidth} %
\epsfig{file=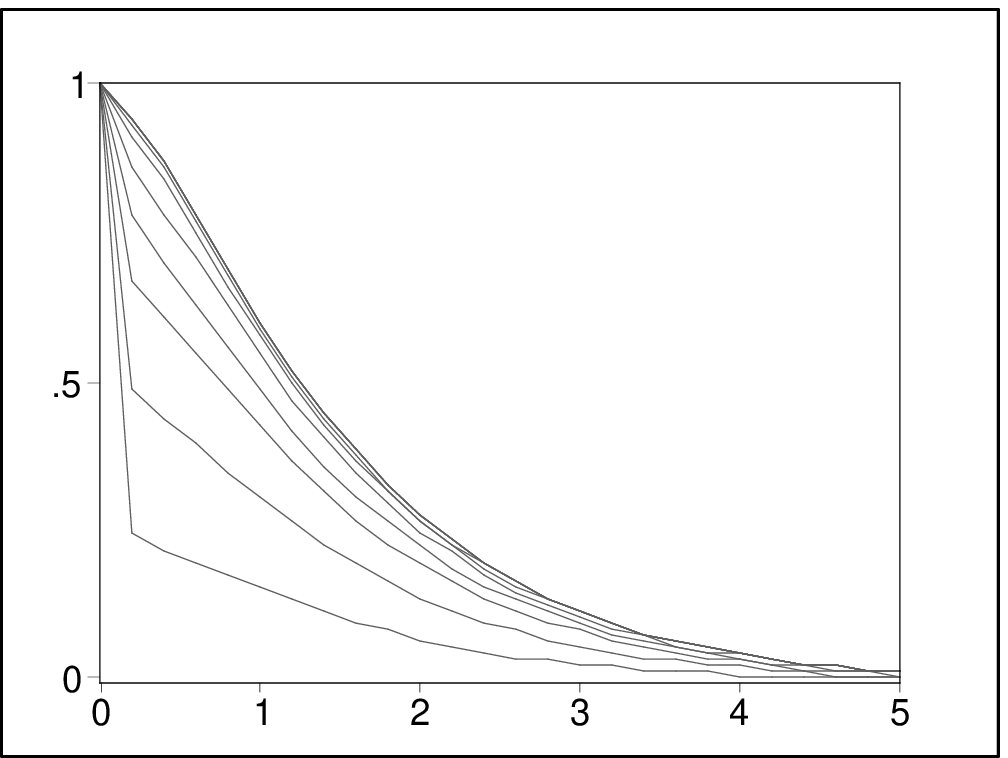,width=0.4\linewidth}
\end{center}
\caption{{}$X(R)$ and $P(R)$ fields of a vortex in dS spacetime with $l=10,$
for the different values of time. Time goes from $-\protect\tau _{\infty }$
to $0,$ from left to right on the curves.}
\label{fig15}
\end{figure}

\begin{figure}[tbp]
\begin{center}
\epsfig{file=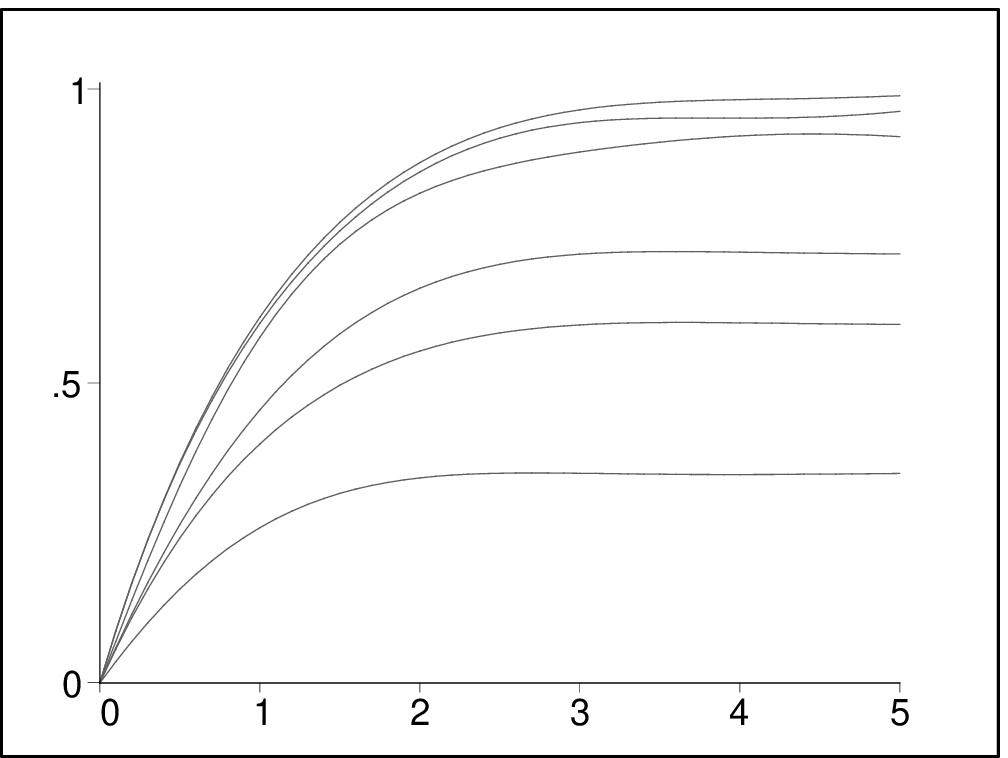,width=0.4\linewidth} %
\epsfig{file=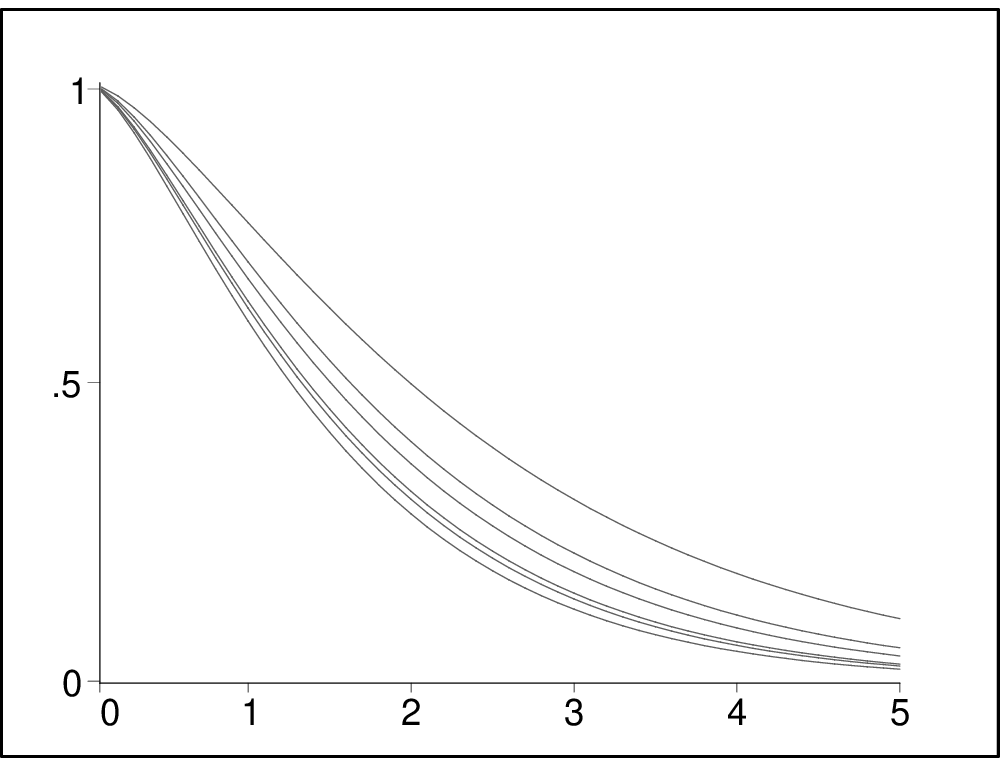,width=0.4\linewidth}
\end{center}
\caption{{}$X(R)$ and $P(R)$ fields of a vortex in dS spacetime with $l=10,$
for the different values of time. Time goes from $0$ to $\protect\tau %
_{\infty },$ from left to right on the curves.}
\label{fig16}
\end{figure}

The preceding calculation of vortex fields on dS spacetimes is not
restricted to the special case of $l=3$; with other values of the
cosmological parameter $l$, we obtain similar results. For example, another
typical results of our calculation are displayed in figures \ref{fig15} and 
\ref{fig16} for value of $\ l=10.$ 

For time slices near $\tau =0,$\ the behaviour of the vortex fields do not
change by changing the cosmological parameter $l.$\ The physical reason is
that at this special time the spacetime is flat and independent of the
cosmological constant, . However, if we consider a constant positive time
slice,\ then by increasing the cosmological parameter $l,$\ the string
thickness decreases; for a constant negative time slice, by increasing the
cosmological parameter $l$, it increases. The asymmetric behaviour of string
thickness for positive and negative times is due to the inflation function $%
e^{2\tau /l}$ in (\ref{bigbang}). Hence by increasing $l$, the density of
vortex field time-constant contours concentrates mainly near the time
constant $\tau =0$\ slice.

To obtain the effect of \ the vortex on the big bang patch of dS$_{4}$\
spacetime, we use the results of the preceding section, in which we found
that vortex induces a deficit angle on the static spacetime metric given in (%
\ref{dsdef}) . By using the following transformations,{\large \ } 
\begin{equation}
\begin{array}{c}
\tau =-t+\frac{l}{2}\ln \left| 1-\frac{r^{2}}{l^{2}}\right| \\ 
X=\frac{r}{\sqrt{\left| 1-\frac{r^{2}}{l^{2}}\right| }}\exp (t/l)\sin \theta
\cos (\alpha \varphi ) \\ 
Y=\frac{-r}{\sqrt{\left| 1-\frac{r^{2}}{l^{2}}\right| }}\exp (t/l)\sin
\theta \sin (\alpha \varphi ) \\ 
Z=\frac{r}{\sqrt{\left| 1-\frac{r^{2}}{l^{2}}\right| }}\exp (t/l)\cos \theta
\end{array}
\label{trans}
\end{equation}
the following metric 
\begin{equation}
ds^{2}=-d\tau ^{2}+e^{2\tau /l}(dX^{2}+dY^{2}+\,dZ^{2})  \label{bigbangdef}
\end{equation}
can be written in the well known static dS spacetime with deficit angle $%
\alpha $\ given by (\ref{dsdef}). Hence the effect of the vortex on a big
bang patch of dS spacetime is to create a deficit angle in the X-Y plane
that is constant as the (locally) flat spatial slice evolves in time.

Similar calculations for larger winding numbers show that increasing the
winding number yields a greater vortex thickness in each constant time\
slice relative to winding number $N=1.$ This tendency runs counter to that
of increasing $l$,\ for which\ the vortex thickness decreases on a constant
positive time slice. As $l\rightarrow \infty $ we can increase the winding
number to larger and larger values, and we find that the resultant thickness
increases. Hence the effect of increasing winding number to thicken the
vortex dominates over the thinning of the vortex due to decreasing
cosmological constant.

What we have found about the behaviour of the vortex in the big bang patch
of dS spacetime can be straightforwardly generalized to the big crunch patch
of the dS spacetime. The big crunch patch is given by replacing $\tau
\rightarrow -\tau $\ in the metric (\ref{bigbang}). Increasing the time from 
$-\tau _{\infty \text{ }}$to $\tau _{\infty },$\ for which the constant time
slices of the big crunch spacetime become smaller and smaller, the string
thickness becomes narrower and narrower. In fact, increasing the time, the
energy density of the string concentrates to smaller and smaller distances
due to the smaller size of the\ constant time slices.

\section{De Sitter $c$-function}

Obtaining evidence in support of (or against) a conjectured dS/CFT
correspondence is somewhat harder to come by than its AdS/CFT counterpart. \
Although it is tempting to think of the former as a `Wick-rotation' of the
latter, a number of subtleties arise whose physical interpretation is not
always straightforward \cite{GH}. \ 

One way of making progress in this area is via consideration of the UV/IR
correspondence. In both the AdS and dS cases there is a natural
correspondence between phenomena occurring near the boundary (or in the deep
interior) of either spacetime and UV (IR) physics in the dual CFT. Solutions
that are asymptotically dS lead to an interpretation in terms of \
renormalization group flows and an associated generalized dS $c$-theorem.
This theorem states that in a contracting patch of dS spacetime, the
renormalization group flows toward the infrared and in an expanding
spacetime, it flows toward the ultraviolet \cite{Leb}. More precisely, the $c
$-function in $(n+1)$-dimensional inflationary dS is given by, 
\begin{equation}
c=\left( G_{\mu \nu }n^{\mu }n^{\nu }\right) ^{-\left( n-1\right) /2}=\frac{1%
}{\varrho ^{\frac{n-1}{2}}}  \label{cfunction}
\end{equation}
where $n^{\mu }$\ is the unit normal vector to a constant time slice and $%
\varrho $ is the energy density on a constant time hypersurface.

It is natural to consider the time-dependent solution of the previous
section in this context. Although we have not solved the
Einstein-Abelian-Higgs equations exactly, we have shown to leading order
that the vortex solution we obtained induces a deficit angle into de Sitter
spacetime, and so to this order our solution is (locally) asymptotically de
Sitter. Furthermore, the gauge field values of our solution asymptotically
approach the constant values expected in a de Sitter vacuum, and so at large
distances (near the boundary) we expect that our solutions will in general
be asymptotically de Sitter, at least locally.

\begin{figure}[tbp]
\begin{center}
\epsfig{file=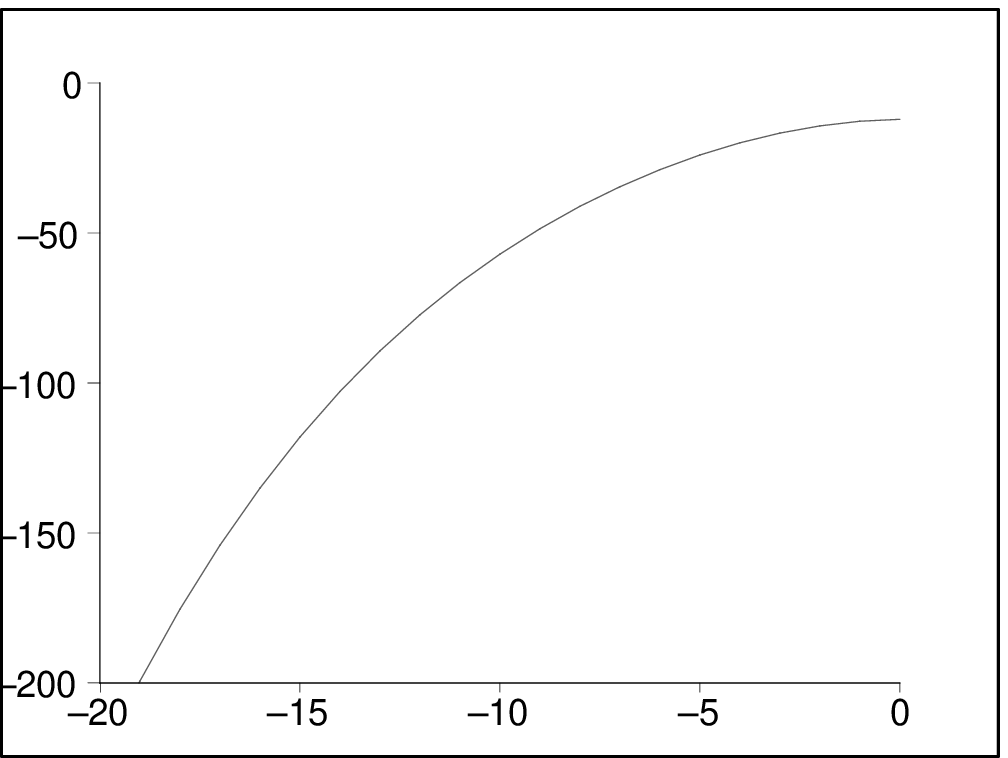,width=0.4\linewidth} %
\epsfig{file=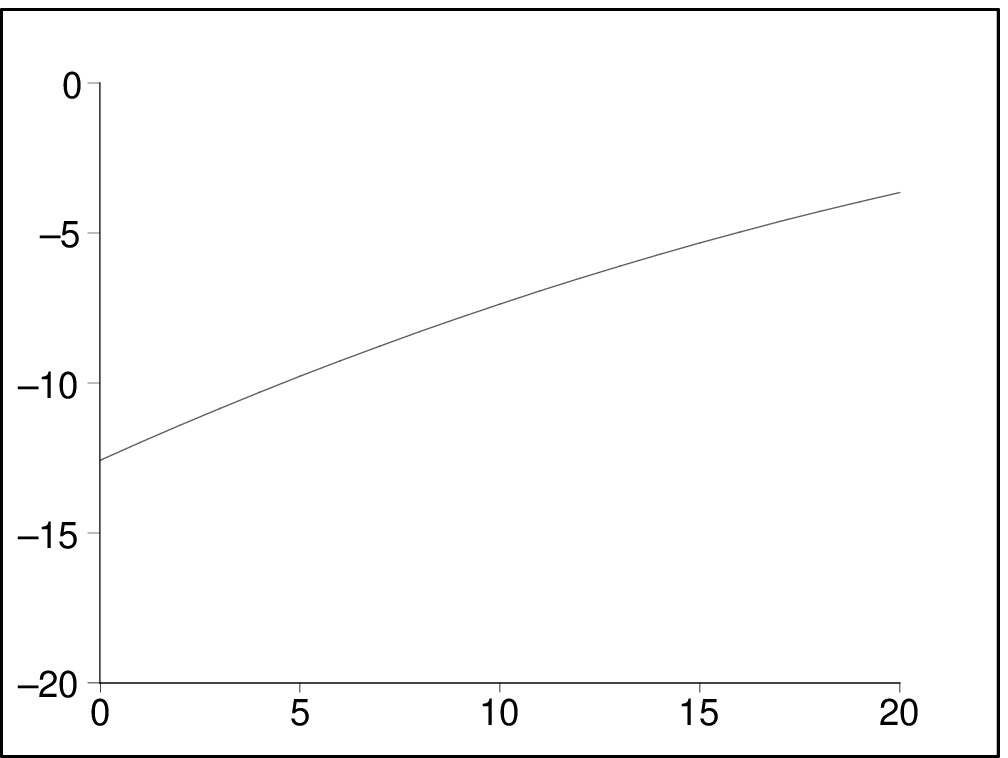,width=0.4\linewidth}
\end{center}
\caption{{}The $c$-function versus time.}
\label{fig17}
\end{figure}

So we want to examine the behaviour of the (\ref{cfunction}) in the big-bang
patch (\ref{bigbang}) in \ the presence of a vortex. The energy density of
the vortex is given by: 
\begin{equation}
\varrho =-\frac{1}{2e^{2\tau }}(\frac{\partial X}{\partial R})^{2}-\frac{1}{%
2R^{2}e^{4\tau }}(\frac{\partial P}{\partial R})^{2}-(X^{2}-1)^{2}-\frac{%
X^{2}P^{2}}{2R^{2}e^{2\tau }}  \label{energyvortex}
\end{equation}
Using the curves of the vortex fields we obtained for different time slices
in figures (\ref{fig13}) and (\ref{fig14}), we get figure (\ref{fig17}) for
the $c$-function in terms of time at a fixed point $R=2$ of space. 

We find that by increasing the time from $-\tau _{\infty }$ to $\tau
_{\infty }$, the $c$-function monotonically increases as the universe
expands. We emphasize that this monotonic increase is not restricted to this
special value of $R$: by changing the value of $R$ we find the same
increasing behaviour of the $c$-function as time evolves.

Similar calculations show that in the big-crunch patch of dS spacetime, the $%
c$-function decreases monotonically as time evolves from $-\tau _{\infty }$
to $\tau _{\infty }.$

\section{Conclusion}

We have solved the Nielsen-Olesen equations in a static dS$_{4}$ background,
and found that the Higgs and gauge fields are axially symmetric, with
non-zero winding number. Our solution in the limit of \ large $l$ (small
cosmological constant) reduces to the well known flat-space solution. The
solution (to leading order in the gravitational coupling) induces a deficit
angle in dS$_{4}$. We find in the static patch that an increasingly positive
cosmological constant tends to make a thicker vortex solution due to
cosmological expansion.

We solved the same equations in the big bang patch of the dS$_{4}$
background, and found that by increasing the time from early times at $\tau
\rightarrow -\infty $ \ to $\tau =0$, the string thickness increases from an
infinitesimally small value to a finite size. Increasing the time to vary
large values $\tau \rightarrow +\infty ,$ the string thickness grows more
and more, so that at future infinity, the string fields approach constant
values everywhere. At past infinity, which is in fact the dS$_{4}$\ horizon,
although the magnitude of the magnetic field goes to infinity near the
string axis, the spatial dimensions shrink relative to spatial dimensions at
other times, yielding a constant value for the magnetic flux of the vortex.
In contrast to past infinity, at future infinity the magnetic field is very
tiny over a big range of \ spatial coordinate, giving rise to the same value
of the magnetic flux of the vortex at past infinity. The presence of a
vortex in asymptotically dS spacetime technically violates the cosmic
no-hair theorem in that the spacetime is not pure de Sitter; rather it is
only locally dS due to the deficit angle induced from the non trivial
topology of the vortex solution. Violation of the cosmic no hair theorem has
also been observed in Einstein-Maxwell-Dilaton theory with a positive
cosmological constant \cite{Ma}.

Our results are in accord with the generalized dS $c$-theorem, providing
further evidence in favor of a conjectured dS/CFT correspondence. An
interesting future challenge is that of obtaining a holographic description
of a vortex solution in dS spacetimes. The boundary field theory will have
to be on a sphere with the same deficit angle as that induced by the vortex.
In the AdS case, the conformal two-point correlation function is obtained by
evaluating the bulk propagator of a scalar field between two points on the
boundary by integrating over all spacelike geodesics paths between the
points, and the presence of a vortex is signified by a discontinuity in this
correlation function \cite{Deh}. However in the dS case the geodesic paths
will be timelike, and will necessarily have to penetrate the cosmological
horizon to detect a vortex in the static patch.\ Not all of \ ${\cal I}^{+}$
( ${\cal I}^{-}$ ) is causally connected\ to the vortex in the big bang (
big crunch ) patch, raising issues of causality reminiscent of those
considered in the case of particles forming black holes in $(2+1)$
dimensions \cite{Lou}, and their resolution is far from clear.

We close by commenting on the physical relevance of our solutions. Since we
employ Planck units $G=\hbar =c=1$ , the constant $l$ is in units of the
Planck length. \ We have found that only for $l\lesssim 10$ do our solutions
appreciably differ from the flat space case. \ In this regime the radius of
curvature of the de\ Sitter spacetime becomes comparable to the Planck
length, equalling it at $l=1$, and\ one might be concerned that our results
will be substantively modified due to quantum gravitational effects. \
Nevertheless we maintain that our results have both utility and validity for
the following reasons. \ First, the qualitative behaviour of our solutions
for large $l$ is the same as for $l\lesssim 10$ ; for example figure (\ref
{fig17}) will retain the same qualitative features for all finite values of $%
l$, and we have numerically checked this for $l=100$. \ However smaller
values of $l$ more clearly highlight the significance of the features of
these solutions, which is why we have presented them here. \ Second, it is
at the very least useful to know what the classical physics is that
underlies any quantum effects, and our solutions furnish will such knowledge
when it becomes available. Finally, quantum gravitational effects will
typically induce quantitative corrections to our results of order $l^{-2}$,
which is still only about $10\%$ even for $l=3$. Hence we expect the
qualitative features of our solutions to remain even when quantum gravity is
taken into account.

Inclusion of quantum gravitational effects is, of course, important. Other
problems include the generalization of our solutions to asymptotically dS
spacetimes with black holes and the possible dS/CFT correspondence of these
solutions. Work on these problems is in progress.

\bigskip {\Large Acknowledgments}

This work was supported by the Natural Sciences and Engineering Research
Council of Canada.\ We would like to thank F. LeBlond for discussions and
comments on an earlier version of this manuscript.

\end{document}